\documentclass[sigconf]{acmart}
\AtBeginDocument{%
  \providecommand\BibTeX{{%
    \normalfont B\kern-0.5em{\scshape i\kern-0.25em b}\kern-0.8em\TeX}}}

\setcopyright{acmcopyright}
\copyrightyear{2018}
\acmYear{2018}
\acmDOI{XXXXXXX.XXXXXXX}

\acmConference[Conference acronym 'XX]{Make sure to enter the correct
  conference title from your rights confirmation emai}{June 03--05,
  2018}{Woodstock, NY}
\acmPrice{15.00}
\acmISBN{978-1-4503-XXXX-X/18/06}

\usepackage{booktabs} 
\usepackage{amsmath}
\usepackage{graphics}
\usepackage{epsfig}
\usepackage{graphicx}
\usepackage[ruled,vlined, linesnumbered]{algorithm2e}
\usepackage{xcolor}
\usepackage[skip=0pt]{caption}
\usepackage{subfigure}
\usepackage{balance}
\usepackage{multirow}
\usepackage{mathrsfs}
\usepackage{acronym}
\usepackage{placeins}
\usepackage{tabularx}
\usepackage{makecell}
\usepackage{xcolor}
\usepackage[inline]{enumitem}
\usepackage{fancyhdr}
\usepackage{soul}

\SetKwInput{KwInput}{Input} 
\SetKwInput{KwOutput}{Output}

\newcommand{\ourmodel}{\textsc{Lite-LLM4Rec}\xspace}

\begin{document}

\title{Rethinking Large Language Model Architectures \\for Sequential Recommendations}


\author{%
Hanbing Wang$^1$\,\,, Xiaorui Liu$^3$\,\,, Wenqi Fan$^4$\,\,, Xiangyu Zhao$^5$\,\,, Venkataramana Kini$^2$ \\ Devendra Yadav$^2$\,\,, Fei Wang$^2$\,\,, Zhen Wen$^2$\,\,, Jiliang Tang$^1$\,\,, Hui Liu$^1$}
\affiliation{%
\institution{$^1$Michigan State University, $^2$Amazon, 
$^3$North Carolina State University  \country{USA}}}

\affiliation{\institution{$^4$The Hong Kong Polytechnic University, $^5$City University of Hong Kong \country{China}}}
\email{{wangh137,liuhui7,tangjili}@msu.edu, {venkini,feiww,yaddevn,zhenwen}@amazon.com}
\email{{liu96@ncsu.edu, wenqifan03@gmail.com, xianzhao@cityu.edu.hk}}
\renewcommand{\shortauthors}{Hanbing et al.}


\begin{abstract}
Recently, sequential recommendation has been adapted to the LLM paradigm to enjoy the power of LLMs.  LLM-based methods usually formulate recommendation information into natural language and the model is trained to predict the next item in an auto-regressive manner. Despite their notable success, the substantial computational overhead of inference poses a significant obstacle to their real-world applicability. In this work, we endeavor to streamline existing LLM-based recommendation models and propose a simple yet highly effective model \ourmodel. The primary goal of \ourmodel is to achieve efficient inference for the sequential recommendation task. 
\ourmodel circumvents the beam search decoding by using a straight item projection head for ranking scores generation. This design stems from our empirical observation that beam search decoding is ultimately unnecessary for sequential recommendations. 
Additionally, \ourmodel introduces a hierarchical LLM structure tailored to efficiently handle the extensive contextual information associated with items, thereby reducing computational overhead while enjoying the capabilities of LLMs.
Experiments on three publicly available datasets corroborate the effectiveness of \ourmodel in both performance and inference efficiency (notably 46.8\% performance improvement and 97.28\% efficiency improvement on ML-1m) over existing LLM-based methods. Our implementations will be open sourced.
\end{abstract}

\ccsdesc[500]{Information systems~Recommender systems}

\keywords{Recommender Systems, Large Language Models, Sequential Recommendation}



\maketitle


\section{Introduction}

Sequential recommendation is to predict next item a user will interact with based on his/her interaction history. Because user interests are dynamic and evolving over time, it is important to capture the sequential pattern, leading to accurate recommendations. Traditional methods model the item transition patterns based on Markov Chain~\cite{he2018translation,rendle2010factorizing,he2016fusing}. With the development of deep learning, a variety of deep neural networks, such as Transformer~\cite{kang2018self,sun2019bert4rec}, RNN~\cite{wu2019session,hidasi2018recurrent} and CNN~\cite{tang2018personalized}, have been proposed to advance the task, achieving remarkable performance. Furthermore, side information (e.g., attributes, titles) has been incorporated~\cite{hidasi2016parallel,huang2019taxonomy,zhang2019feature}, which helps achieve remarkable improvement, demonstrating its importance and potential. 

\begin{figure}
    \centering
    \includegraphics[scale=0.65]{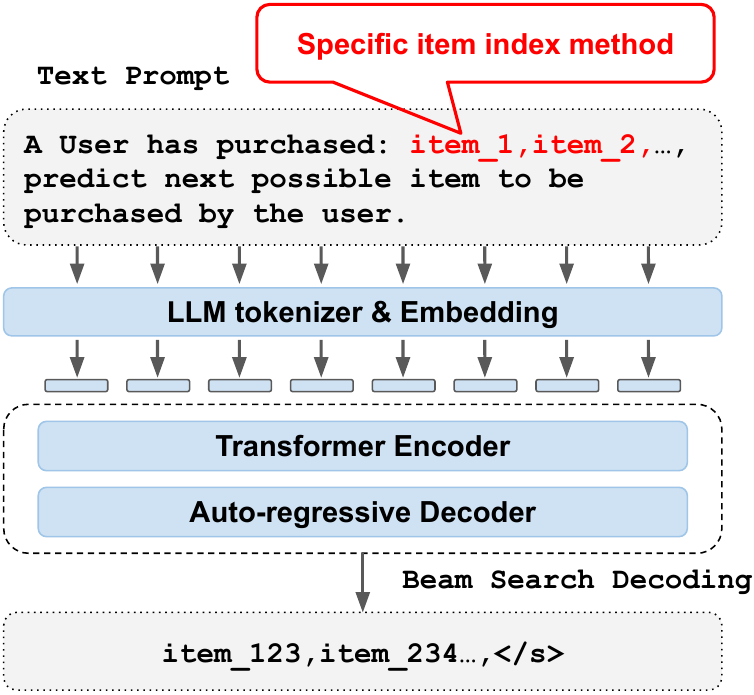}
    \caption{An Illustration of LLM-based sequential recommendations.}
    \label{fig:commonLLM}
\end{figure}

Recently, the widespread success of large language models (LLMs), such as GPT~\cite{brown2020language}, T5~\cite{raffel2020exploring}, and Llama~\cite{touvron2023llama}, has demonstrated their exceptional ability of contextual understanding and offers a promising direction to improve recommendation systems with heightened personalization and adaptability. 
Existing LLM-based recommendation algorithms~\cite{geng2022recommendation,zhang2023recommendation,li2023prompt} mainly adapt recommendation tasks to the LLM paradigm by formulating relevant information, i.e., interaction information, meta data, or candidate items through various indexing strategies into natural language. As shown in Figure~\ref{fig:commonLLM}, such information will be wrapped in a prompt template, and fed into LLMs as input. 
Subsequently, the input will be transformed into an informative latent representation and the model will auto-regressively generate the recommendations in natural language through decoding with beam search~\cite{yang2023large}. 

Although existing LLM-based methods have achieved remarkable success, the exorbitant cost of inference hinders their real-world applications and poses a formidable obstacle to seamless, real-time user experiences~\cite{rajput2023recommender,yue2023llamarec,mei2023lightlm}.  To gain deeper insights into this efficiency challenge, we perform a preliminary study in Section~\ref{sec:3.2} on the impact of the key components of LLM-based methods. First, we observe that the beam search decoding is the most time-consuming component. This component is employed to generate $k$ recommendations for a user, leading to $k$ times greater model computation complexity. Second, we find that inference time diminishes with reduced input length. Longer input often leads to more model computations~\cite{li2023prompt}. Furthermore, we reveal that existing item indexing will lead to serious redundant computation problem because 1) the item indexing will be tokenized into several tokens before being fed into the LLM, which will result in longer input; and 2) some items may appear frequently in the data and the model will compute on the item token string every time when it appears. Our further study of the impact of beam search and item indexing on model performance in Section~\ref{sec:3.3} suggests that the beam search decoding and item indexing are burdensome and unnecessary for sequential recommendation. 

Grounded on our preliminary studies, we aim to streamline the model architecture of existing LLM-based recommendations, introducing \ourmodel as an efficient solution for sequential recommendation. 
We find that the beam search decoding is unnecessary and resource-intensive for this specific task. 
Therefore, \ourmodel circumvents the beam search decoding process. 
Though representing items with their context information can achieve superior performance, it makes the input very long and correspondingly increases computational costs. To tackle this challenge, we propose a hierarchical LLM structure with two key components - Item LLM and Recommendation LLM - to process extensive context information efficiently.
We demonstrate that formulating
relevant information such as interaction information, meta information into natural language is superfluous because LLMs can actually understand the semantic information encoded in the latent representation. 
This design can reduce the input length and alleviate redundant computation while enjoying the power of LLMs. 
Experimental results indicate that \ourmodel can significantly improve not only the inference efficiency but also the overall performance. 
The main contributions of this paper are summarized as follows: 
\begin{itemize}
    \item We simplify existing LLM-based recommendation models and propose a simple but effective sequential recommendation model \ourmodel. Through eliminating the decoder, \ourmodel can achieve better performance as well as significant inference efficiency.
    \item We design a novel hierarchical LLM structure to efficiently process long context information in LLM-based recommendations, which can reduce the computational demands while enjoying the power of LLMs.
    \item We demonstrate the effectiveness of \ourmodel on various benchmark datasets. 
\end{itemize}

\section{Related work}

In this section, we summarize relevant works on sequential recommendation and LLM-based solutions.

\subsection{Sequential Recommendation}
Sequential recommendations~\cite{kang2018self, sun2019bert4rec} leverage user historical interactions to infer the next item the user will interact with. Since user interests are dynamic and evolving over time, it is important to capture this sequential pattern and provide appropriate recommendations. Early studies mainly depend on Markov Chain to model item transition patterns~\cite{rendle2010factorizing, he2016fusing}. Recently, deep-learning based methods have dominated the area. 
For example, GRU4rec~\cite{hidasi2015session} proposes to use RNN in session-based recommendation. Some approaches introduce CNN into sequential recommendation~\cite{tang2018personalized, yan2019cosrec}.
SASrec and BERT4rec~\cite{kang2018self, sun2019bert4rec} employ the self-attention mechanism into sequential recommendation and achieve excellent performance. However, these methods neglect rich contextual information about items, which is important for item modeling. To tackle this problem, several algorithms are proposed~\cite{yuan2020attention,liu2016context, manotumruksa2018contextual,zhang2019feature,zhou2020s3}. For example, FDSA~\cite{zhang2019feature} proposes a self-attention block to leverage item attribute information. $S^3$rec~\cite{zhou2020s3} maximizes the mutual information of context information in different forms to improve sequential recommendation. 
Despite the remarkable improvements made by these methods, they can still be further improved via the excellent world knowledge and context understanding ability of LLMs.

\subsection{LLM-based Recommendation}
The prevalence of LLMs has introduced a paradigm shift into recommender systems~\cite{fan2023recommender}.
Early approaches investigate the practicality of textual representations generated by a language model for recommendations ~\cite{yuan2023go,li2023exploring,qiu2021u,harte2023leveraging}.  
By pre-training and then fine-tuning on downstream datasets, enhanced representations can be obtained ~\cite{li2023text,ding2021zero, hou2022towards}. 
The emergence of generative LLMs shifts recommendation system towards generative paradigm~\cite{rajput2023recommender}. Early attempts explore the potential of LLMs on recommendation via prompt or in-context learning ~\cite{sun2023chatgpt,liu2023chatgpt,thomas2023large}.
TallRec~\cite{bao2023tallrec} trains the LLM to predict whether a user will like a new item given users' interaction history, underscoring the importance of instruction tuning.
P5~\cite{geng2022recommendation} reformulates several recommendation tasks into a natural language generation task via personalized prompts. 
Refer to Figure ~\ref{fig:commonLLM} for an overall understanding of its architecture. 

In order to leverage the power of LLMs, items are usually represented in natural language form via specific item indexing methods, which can be roughly divided into two categories. The first category proposes to reflect item relations via share tokens, such as semantic IDs ~\cite{hua2023index,mei2023lightlm}.
Given the continued significance of ID information, recent studies keep the IDs and represent items as 'item\_1234', which will be tokenized into a token sequence 'item','\_','12','34' before being input to LLMs~\cite{li2023prompt,qiu2023controlrec,geng2022recommendation}. The second category represents items via context information. Tallrec ~\cite{bao2023tallrec} uses item title to represent items. RECFORMER~\cite{li2023text} flattens the key-value attribute pairs of the item as a sequence to accommodate more textual information, and employs a Longformer~\cite{beltagy2020longformer} to process the long context. 

Although these methods have achieved remarkable improvements, they still face slow inference problem. Recent studies have delved into this issue. POD~\cite{li2023prompt} improves P5~\cite{geng2022recommendation} by distilling discrete prompt into continuous prompt to reduce the input length, thus reducing inference time. llamaRec~\cite{yue2023llamarec} proposes a two stage framework. Specifically, it retrieves candidate items via traditional ID-based methods, and designs a verbalizer approach for re-ranking. E4SRec~\cite{li2023e4srec} proposes to integrate ID embeddings extracted from a pre-trained SASRec with instruction prompt but neglects the rich context information of items. LightLM~\cite{mei2023lightlm} proposes a tailored transformer-based architecture to achieve effective recommendations. It improves the inference efficiency by reducing the number of neurons thus reducing the computation demands. However, the modification of architecture may destroy the pre-trained knowledge of LLMs, leading to sub-optimal performance. Despite the strides made in improving inference efficiency, these methods still fall short in addressing the most time-consuming component and the issue of redundant computation.


\section{Preliminaries}
\label{sec:Preliminaries}

In this section, we empirically study the inference efficiency problem of existing LLM-based recommendations. First, we compare the inference time of LLM-based methods with traditional sequential recommendation algorithm BERT4rec ~\cite{sun2019bert4rec} and analyse the impact of each component in LLM-based recommendations in terms of inference efficiency. Then, we investigate the impact of different components in terms of performance effectiveness. Before detailing the aforementioned preliminary studies, we first introduce the problem statement and some key notations we will use in this work.

\subsection{Problem Statement and Notations}
 We denote $u \in \mathcal{U}$ and $i \in \mathcal{I}$ for a user and an item, where $\mathcal{U}$ and $\mathcal{I}$ indicate the user set and the item set, respectively. The interaction history of a user $u$ can be organized as a sequence $\mathcal{I}_u=(i_1,i_2,...,i_t)$ in a chronological order, where $t$ is the length of $\mathcal{I}_u$ and each item $i$ is associated with textual information $\mathcal{T}_i$ (e.g., title and genre). 
 Given the interaction history $\mathcal{I}_u$ of user $u$, sequential recommendation algorithms aim to predict next item $i_{t+1}$ the user is most likely to interact with from $\mathcal{I}\backslash \mathcal{I}_u$, which represents the item set formed by excluding the items already interacted with the user from the complete set of items $\mathcal{I}$.

 \begin{figure}
    \centering
    \includegraphics[clip,trim=2mm 0mm 0mm 0mm,width=0.5\textwidth]{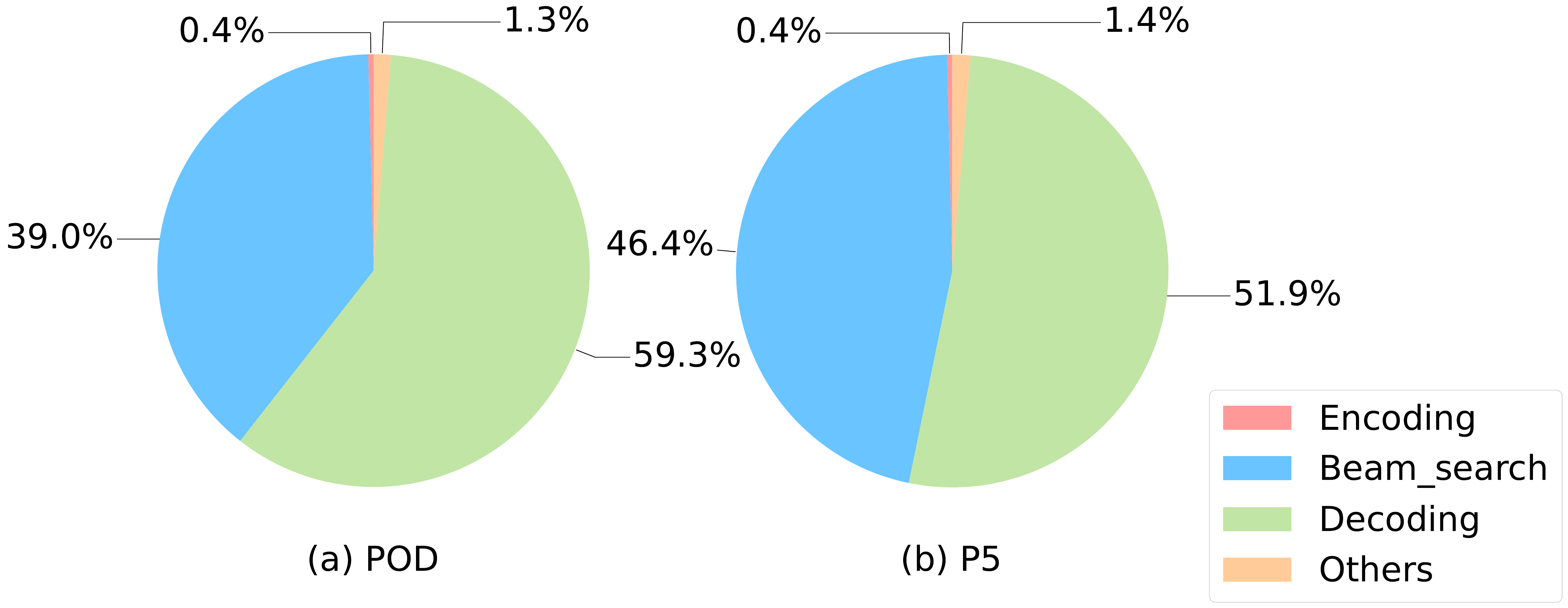}
    \caption{Inference time of different components for a batch of 32 users. 'Beam\_search' refers to the beam search decoding process, 'Decoding' transfers the latent representation to token IDs, and 'Others' includes data preparation, metrics calculations etc.
    }
    \label{fig:time_components}
    \vspace{0.1cm}
\end{figure}

\subsection{Efficiency Analysis}
\label{sec:3.2}

\begin{table}[t]
\caption{Inference time (ms) and input length}
\label{tab:intro1}
\centering
    \begin{tabular}{l rrrr}
    \toprule
    Method & Avg length & Time/Batch \\ 
    \midrule
    Bert4rec  & 21.0  & 2.37  \\
    \midrule
    P5  & 112.2  & 2280 \\
    POD  & 85.0  & 2170  \\
    \bottomrule
    \end{tabular}
\end{table}

In this subsection, we will explore the following questions: 1) In terms of inference efficiency, how do LLM-based recommendation algorithms perform compared with traditional recommendations; and 2) Why do they face such serious inference efficiency problem. To answer these questions, we first demonstrate the huge inference efficiency gap between LLM-based and traditional sequential recommendation systems and then investigate the inference time cost of each component, and the relation between input length and inference time. In this study, we focus on two typical LLM-based recommendation algorithms, i.e., P5 and POD~\cite{geng2022recommendation,li2023prompt}, and one representative traditional sequential recommendation system, i.e., BERT4rec~\cite{sun2019bert4rec}. 

For all the following experiments, we use the implementations released by POD~\cite{li2023prompt}\footnote{\url{https://github.com/lileipisces/POD}} and we keep the same experimental settings.
Our analysis is based on ML-1m and Movies, two popular publicly-available datasets. More details of the datasets can be found in Section~\ref{sec:data}. We calculate the inference time by measuring the total time of all the test data going through each component during inference, and then we can obtain the average inference time for a batch of 32 users. We denote the length of the input as the number of tokens of the tokenized context (e.g., ’item\_1234' will be tokenized as ’item’ ’\_’ ’12' '34' by the tokenizer of T5, and the input length will be 4). The results are shown in Table~\ref{tab:intro1}. We can make the following observations: 1) Compared with Bert4rec, the efficiency of P5 and POD has been significantly compromised, experiencing a slowdown of nearly a thousand times; and 2) POD is more efficient than P5 via reducing the input length which can lead to the decrease of the computational costs. To further explore why LLM-based methods face such formidable inference problem, we conduct time analysis on the inference time cost of each component in P5 and POD~\cite{geng2022recommendation, li2023prompt}.
Figure~\ref{fig:time_components} demonstrates the average inference time cost of each component. From the figure,  we note that the beam search decoding process is most time-consuming which takes approximately 98.2\% on P5~\cite{geng2022recommendation} and 98.3\% on POD~\cite{li2023prompt} of the inference time.

\begin{table}[t]
\caption{Performance of P5 and POD and their two variants}
\label{tab:preP5}
\centering
    \begin{tabular}{l rrrrrr}
    \toprule
    \multicolumn{1}{l|}{Datasets} & \multicolumn{3}{c|}{ML-1m} & \multicolumn{3}{c}{Movies} \\
    \multicolumn{1}{l|}{Methods} & R@20 & N@20 & \multicolumn{1}{l|}{Time} & R@20 & N@20 & Time\\ 
    \midrule
    \multicolumn{1}{l|}{P5}  & 0.2985  & 0.1442 & \multicolumn{1}{l|}{2,280} & 0.1080 & 0.0761 & 1,770 \\
    \multicolumn{1}{l|}{w/o\_d}  & 0.3109
  & 0.1459 & \multicolumn{1}{l|}{40.06} & 0.1217 & 0.0794 & 37.01 \\
    \multicolumn{1}{l|}{w/o\_d\_TID}  & 0.3354 & 0.1586 & \multicolumn{1}{l|}{95.40} & 0.1401 & 0.0905 & 99.02 \\
    \midrule
    \multicolumn{1}{l|}{POD}  & 0.2992  & 0.1403 & \multicolumn{1}{l|}{2,170} & 0.1089 & 0.0761 & 1,400  \\
    \multicolumn{1}{l|}{w/o\_d}  & 0.3022  & 0.1416 & \multicolumn{1}{l|}{35.21} & 0.1330 & 0.0866 & 32.13\\
    \multicolumn{1}{l|}{w/o\_d\_TID}  & 0.3339 & 0.1567 & \multicolumn{1}{l|}{90.26} & 0.1406 & 0.0908 & 80.38\\
    
    \bottomrule
    \end{tabular}
\end{table}

\subsection{Effectiveness Analysis}
\label{sec:3.3}
In the previous subsection, we have demonstrated that the inefficiency of LLM-based recommendations comes from the beam search decoding process and input length. 
In this subsection, we investigate how the beam search decoding and item indexing affect the inference efficiency and the performance. We implement two variants of P5 and POD as follows:
\begin{enumerate}
    \item w/o\_d. It eliminates the decoder and uses an item projection head to perform the recommendation task (Details in Section \ref{itemprojectionhead}).
    \item w/o\_d\_TID. On the basis of eliminating the decoder, it represents items with their titles instead of random numbers. 
\end{enumerate}

We report the performance and inference efficiency in Table ~\ref{tab:preP5}.
We can make the following observations: 1) eliminating the beam search decoding can significantly improve the inference efficiency; 2) although representing items with their titles can improve the performance due to the incorporation of contextual information, it impairs the inference efficiency because the length of input becomes longer (increases to 297.4 on ML-1m), resulting in more computational costs.

\subsection{Discussion}
In this subsection, we summarize key findings from the preliminary studies as follows: 
(1) Compared with Bert4Rec, LLM-based recommendations are much more time-consuming. 
(2) Reducing the input length will improve the inference efficiency. 
(3) Beam search decoding will have negative impacts on the efficiency of sequential recommendation.
(4) Compared with random number, item title can better represent items and achieve better performance due to the incorporation of contextual information. 
(5) Despite the advantages of item title, they are usually very long and will increase computational costs. These findings provide the groundwork for us to simplify existing LLM-based recommendations and propose a simple but effective framework \ourmodel for sequential recommendations. 

\section{The Proposed Framework}

Motivated by our findings, in this section, we aim to simplify the architecture to obtain a better sequential recommendation system, which is easier to train and can achieve low-latency inference and better performance. In this section, we introduce the proposed framework \ourmodel. We first give an overview of \ourmodel. Then we detail its key components and finally give its training details.

\subsection{An Overview}

\begin{figure}
    \centering    \includegraphics[scale=0.7]{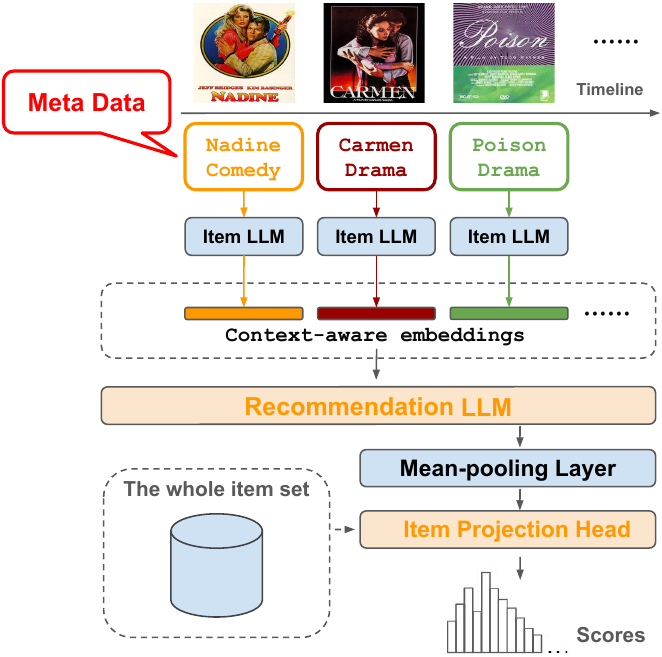}
    \caption{An overview of the architecture of \ourmodel.
    }
    \label{fig:model}
    \vspace{0.2cm}
\end{figure}

Figure~\ref{fig:model} demonstrates the whole architecture of \ourmodel. To mitigate redundant computation and enhance inference efficiency, \ourmodel proposes a hierarchical LLM structure which contains two LLM components: Item LLM and Recommendation LLM. We first encode the extensive context information of items via the Item LLM into context-aware vectors and then the Recommendation LLM takes these context-aware vectors as input instead of the original lengthy contexts.
Then, \ourmodel proposes to replace the beam search decoding process with an item projection head. As a result, it can generate recommendations more effectively. In the following subsections, we will detail our model.

\subsection{Hierarchical LLM Structure}
In order to leverage the power of LLMs, existing LLM-based recommendations usually formulate items into natural language through various indexing strategies, which can be fed into LLMs. These item indexing methods can be roughly divided into two categories. The first category encodes item relations into the item indexing such as semantic IDs~\cite{hua2023index,li2023prompt,mei2023lightlm}. These methods can capture item relations via shared tokens but the indexing by itself doesn't contain any semantic information, which may not be able to fully explore the potential of LLMs. The second category denotes an item by its meta data such as title or genre~\cite{bao2023tallrec,li2023text}. These methods take item contextual information into consideration but the inputs are usually very long, which needs more time to process or even worse they need truncation or special architectures to process when they exceed the length limitation of LLMs. 

In addition, the above two kinds of item indexings will be tokenized before being fed into the LLM. This tokenization process leads to increased computational complexity, as each item will be represented by multiple tokens.
Moreover, redundant computations occur when the same item appears multiple times in the input. 
For example, in the ML-1m dataset, the movie 'Star Wars: Episode I - The Phantom Menace (1999)' will be tokenized into 11 tokens by T5~\cite{raffel2020exploring}, and appears 539 times in the input during inference, leading to severe redundant computation problem and correspondingly hurting the inference efficiency. . 

\ourmodel aims to simplify existing item indexing methods
while also being able to leverage the power of LLMs. Particularly, we propose a hierarchical LLM structure comprising two distinct LLM components:
Item LLM and Recommendation LLM.
The Item LLM is to encode extensive context information of an item into a compact, context-aware vector representation. Leveraging its capabilities, the Item LLM can effectively capture the contextual nuances and dependencies within the input sequence, facilitating the creation of a context-aware vector for each item.
Then, the Recommendation LLM processes the sequence of context-aware vectors rather than the original lengthy context sequences. Thus the input length of LLM can be significantly reduced. 
In the following, we will give the details about acquiring context-aware vectors for items and sequence representations.

We represent the context information of item $i$ (e.g., title, genre) after tokenization by $\mathcal{T}_i^e=(w_1, w_2,...,w_L)$, where $w$ denotes tokens of the context information and $L$ is the length of the context. Then, we input $\mathcal{T}_i^e$ into the item LLM which can be denoted as:
\begin{equation} \label{eq:1}
    Input(\mathcal{T}_i^e) = (w_1, w_2,...,w_L). 
\end{equation}
Through the model, we can obtain the output representation for each token:
\begin{equation} \label{eq:2}
    Output(\mathcal{T}_i^e) = (h_{w_1}, h_{w_2},...,h_{w_L}). 
\end{equation}
Then we can obtain the context-aware embedding for the item:
\begin{equation} \label{eq:3}
    h_i = Mean\_pooling(h_{w_1}, h_{w_2},...,h_{w_L}), 
\end{equation}
where $h_{w_i} \in \mathbb{R}^{1 \times d}$ is the representation for the corresponding token $w_i$. Notice that deriving high quality item representation is not our focus, so we just apply a simple mean-pooling over all tokens' representation to get the final context-aware embedding for each item.
More complicated representation methods will be explored as one future work. 

After obtaining context-aware embeddings for items. The Recommendation LLM will take a sequence of context-aware embeddings as input instead of the lengthy natural language input. Similarly, we also apply a simple mean-pooling function to the output of the Recommendation LLM to obtain the representation of sequence:
\begin{equation} \label{eq:4}
 h_{u}=Mean\_pooling(Rec\_LLM(h_1, h_2, ..., h_t)), 
\end{equation}
where $h$ represents the context-aware embedding for an item, $Rec\_LLM$ indicates the Recommendation LLM with per-trained weights.

This kind of architecture has several advantages: First, we encode the long context information of an item via the Item LLM into a single embedding, which greatly reduces the length of the input sequence while preserving the semantic information. Second, it mitigates the redundant computation problem since the item representation can be directly obtained from the Item LLM instead of computing repetitively every time when the item appears. Thus, \ourmodel can improve inference efficiency via reducing computational costs.

\begin{algorithm}[htb]
\SetAlgoLined
\KwInput{The sequential data $\mathcal{D}$, Context information of items $\mathcal{T}$, hyper-parameter settings;}
\KwOutput{Model parameters $\theta$;}
\While{not coverage}{
    Draw a batch of data $\mathcal{B}$ from $\mathcal{D}$\;
    \For{$(u, \mathcal{I}_u)$ in $\mathcal{B}$}{
    Sample an item segment ${I}_u^{seg}$ from $\mathcal{I}_u^{|s_u|-2}$\;
    \For{$i$ in ${I}_u^{seg}$}{
    Acquire context information $\mathcal{T}_i$\;
    Calculate context-aware embeddings $h_i$ through Eq.~\ref{eq:1} - Eq.~\ref{eq:3}\;
    }
    Acquire the user representation  $h_{u}$ through Eq.~\ref{eq:4}\;
    }
    Compute logits through Eq.~\ref{eq:5} \;
    Compute $L_{CE}$ through Eq.~\ref{eq:6}\;
    Update the parameters $\theta$ of the model through $L_{CE}$ \; 

}
\caption{Training Process of \ourmodel}
\label{training}
\end{algorithm}

\subsection{Item Projection Head}
\label{itemprojectionhead}
Our findings in preliminary studies suggest that the beam search decoding during the auto-regressive generation is the most time-consuming component. This operation is originally designed for natural language generation, which generates the tokens step by step. When it is applied to sequential recommendations, it will generate the recommendation items in natural languages form (e.g., generating a string 'item\_1123'). 

Our preliminary study suggests that this operation is unnecessary for sequential recommendations due to the following disadvantages. First, the head layer of LLM will output the probability distribution over the LLM vocabulary, which is unnecessary since item tokens only account for a small proportion of the whole vocabulary. Second, The model will generate non-existent and repetitive items. Third, generating $k$ recommendations for a user will result in $k$ times more computational costs which significantly contributes to inference delays, thereby exacerbating the efficiency problem.

\ourmodel proposes to circumvent the beam search decoding process and introduces a simple but effective item projection head, which will directly output the probability over the whole item set. \ourmodel treats the probability as the ranking scores and takes items with the highest scores as the final recommendations. In particular, \ourmodel adopts a simple one-layer Multi Layer Perceptron (MLP) without bias ~\cite{kang2023llms} as the projection head, which can be formulated as follows:
\begin{align}
\label{eq:5}
   & logits=W_{proj}h_{u},
\end{align}
where $h_{u}$ stands for sequence representation for user $u$ obtained by Eq.~\ref{eq:4},
$W_{proj}$ is the projection matrix of the MLP, and $logits$ represents the output scores over the whole item set.

\subsection{Model Training}
 
Since we use an item projection head to acquire recommendation scores, there is no need to do negative-sampling because it implicitly considers all the un-interacted items as negative samples. 
We train the hierarchical model with a cross-entropy loss as shown below:
\begin{equation} \label{eq:6}
    L_{CE} = -\sum_{i=1}^Ny_ilogr_i, 
\end{equation}
where $N$ is the number of items, $y_i$ represents the ground-truth for item $i$, which is 1 if item $i$ is the ground-truth
item, otherwise 0; and $r_i$ is the predicted score of item $i$. We provide the training procedure as well as the implementation details in Algorithm~\ref{training}.
We adopt two training strategies which are denoted as '\ourmodel\_sampling' and '\ourmodel\_all' in Table~\ref{tb:main_result} and their details are given in ~\ref{sec:implementation_details}. 

\begin{table}[t]
\caption{Statistics of the datasets after pre-processing.}
\label{table2}
\centering
    \begin{tabular}{l rrrr}
    \toprule
    Dataset & Users  & Items & Interact & Sparsity(\%)\\ 
    \midrule
    ML-1M  & 6,040  & 3,706 & 994,169   & 0.0446   \\
    Movies  & 79,276  & 29,946 & 1,775,011   & 0.0007   \\
    Toys    & 11,803 & 8,569 & 206,103   & 0.0020    \\ 
    \bottomrule
    \end{tabular}
    \vspace{0.2cm}
\end{table}

\begin{table*}[h]
\caption{Overall performance comparison of on ML-1m, Movies and Toys with conventional sequential baselines and LLM-based sequential methods.
R denotes Recall, and N denotes NDCG. Boldface represents the best results. Underscore indicates the second best results.}
\label{tb:main_result}
\centering
    \resizebox{\linewidth}{!}{
        \begin{tabular}{l cccc cccc cccc}
    \toprule
    \textbf{Datasets} & \multicolumn{4}{c}{\textbf{ML-1m}} & \multicolumn{4}{c}{\textbf{Movies}} & \multicolumn{4}{c}{\textbf{Toys}}                               \\ 
    \cmidrule(r){2-5}\cmidrule(r){6-9}\cmidrule{10-13}
    Method            & R@10   & R@20   & N@10   & N@20   & R@10   & R@20   & N@10   & N@20   & R@10   & R@20   & N@10   & N@20   \\ 
    \midrule
    BERT4Rec                & 0.1981 & 0.2892 & 0.1093 & 0.1323 & 0.0761 & 0.1048 & 0.0493 & 0.0565 & 0.0352 & 0.0543 & 0.0179 & 0.0227 \\
    SASRec                & 0.1834 & 0.2785 & 0.0801 & 0.1040 & 0.1082 & 0.1449 & 0.0714 & 0.0807 & 0.0561 & 0.0776 & 0.0312 & 0.0366 \\
    $S^3$-Rec          & 0.1776 & 0.2733 & 0.0882 & 0.1123 & 0.0764 & 0.1074 & 0.0481 & 0.0560 & 0.0538 & 0.0811 & 0.0276 & 0.0345 \\
    FDSA          & 0.1962 & 0.2854 & 0.1058 & 0.1283 & 0.1151 & \ul{0.1500} & 0.0804 & 0.0891 & 0.0568 & 0.0860 & 0.0344 & 0.0417 \\
    LightLM          & 0.1705 & 0.2531 & 0.0928 & 0.1135 & 0.0751 & 0.0958 & 0.0521 & 0.0574 & 0.0509 & 0.0721 & 0.0302 & 0.0355 \\
    P5             & 0.2149 & 0.2985 & 0.1232 & 0.1442 & 0.0905 & 0.1080 & 0.0718 & 0.0761 & 0.0543 & 0.0661 & 0.0356 & 0.0416 \\
    POD             & 0.2185 & 0.2992 & 0.1201 & 0.1403 & 0.0904 & 0.1089 & 0.0715 & 0.0761 & 0.0565 & 0.0690 & \ul{0.0421} & \ul{0.0452} \\
    \midrule 
    \ourmodel\_sampling   & \ul{0.2733} & \ul{0.3770} & \ul{0.1518} & \ul{0.1780} & \ul{0.1156}& 0.1447 & \ul{0.0856} & \ul{0.0929} & \textbf{0.0682} & \textbf{0.0927} & \textbf{0.0426} & \textbf{0.0488} \\
    \ourmodel\_all & \textbf{0.3209} & \textbf{0.4255} & \textbf{0.1866} & \textbf{0.2129} & \textbf{0.1253} & \textbf{0.1596} & \textbf{0.0906} & \textbf{0.0992} & \ul{0.0627} & \ul{0.0894} & 0.0382 & 0.0449 \\
    
    \bottomrule
    \end{tabular}
    }
\vspace{0.2cm}
\end{table*}

\section{Experiment}
\label{sec:exp}
In this section, we conduct comprehensive experiments to verify the effectiveness and efficiency of the proposed \ourmodel. In particular, we try to answer the following questions:
\begin{itemize}
    \item Can the proposed \ourmodel achieve better overall performance? (Section~\ref{sec:5.2})
    \item Can the proposed \ourmodel improve inference efficiency? (Section~\ref{sec:5.3})
    \item How does \ourmodel perform on Top-N recommendation task? (Section~\ref{sec:5.4})
    \item How do different components of \ourmodel affect the recommendation performance? (Section~\ref{sec:5.5})
\end{itemize}

\subsection{Experimental Settings}
\label{sec:expset}

\subsubsection{Datasets}
\label{sec:data}

To evaluate the effectiveness of \ourmodel, we conduct a series of experiments on three real-world benchmark datasets, including ML-1m\footnote{\url{https://grouplens.org/datasets/movielens/1m/}} ~\cite{harper2015movielens}, Amazon-Movies and Amazon-Goys\&Games\footnote{\url{https://cseweb.ucsd.edu/~jmcauley/datasets.html\#amazon_reviews}}~\cite{ni2019justifying}. We partition them into training, validation and test sets with the commonly used leave-one-out strategy. It takes the second-to-last item as the validation item, the last item as the test item and all other items as training items in each user's interaction history. Table~\ref{table2} shows the statistics of these datasets and some details are as follows:

\begin{itemize}
    \item The ML-1m dataset is an open dataset for movie recommendations. There are approximately 100k interactions. We adopt 5-core filtering strategy where we filter out users and items with less than 5 interactions. 
    \item We consider two categories of Amazon dataset corpora: Movies and Toys\&Games (denoted as Toys for clarity). These datasets are collected from the e-commerce platform Amazon\footnote{\url{https://www.amazon.com/}} with item meta data, user reviews and ratings. We adopt 10-core filtering strategy to filter the users and items with less than 10 interactions to ensure data quality.
\end{itemize}

\subsubsection{Evaluation protocols}
We adopt two widely used metrics Recall@$k$ and NDCG@$k$, where $k=10,20$. Recall@$k$ represents the coverage of ground-truth items that appear in the final recommendation list.  NDCG@$k$ measures the ranking quality of the final recommendation items. For both metrics, a larger value indicates better performance. For our method and the baselines, we evaluate the performance on the whole item set, and the reported results are the average values over all users.

\subsubsection{Baselines}

We choose representative methods from three groups as baselines, i.e., traditional ID-based sequential models, Context-aware ID-based models, LLM-based recommendation models. We consider the following tradition ID-based sequential models:
\begin{enumerate*}[label=(\roman*), leftmargin=*]

\item \textbf{SASRec}~\cite{kang2018self} proposes an unidirectional attention-based sequential model which can capture long-term semantics to predict the next item. 

\item \textbf{BERT4Rec}~\cite{sun2019bert4rec} introduces a bidirectional attention-based transformer to model user behavior sequences. It introduces the Cloze objective into sequential recommendations. 
\end{enumerate*}

The context-aware ID-based sequential models include:
\begin{enumerate*}[label=(\roman*), leftmargin=*]

\item \textbf{S3Rec}~\cite{zhou2020s3} applies self-supervised learning to the sequential recommendation task. It proposes four self-supervised optimization objectives to maximize the mutual information of context information to learn the correlation between items.

\item \textbf{FDSA}
~\cite{zhang2019feature} proposes to model feature transitions through different self-attention blocks. It integrates with item-level transitions for modeling user's sequential intents.
\end{enumerate*}

We choose the following LLM-based sequential models:
\begin{enumerate*}[label=(\roman*), leftmargin=*]

\item \textbf{P5}~\cite{geng2022recommendation} transforms various recommendation tasks into the conditional natural language generation task via personalized prompts and integrates them into a unified framework.

\item \textbf{POD}~\cite{li2023prompt} proposes to distill knowledge in the discrete prompt into continuous prompt vectors, which is more flexible and expressive and can reduce the inference time.

\item \textbf{LightLM}~\cite{mei2023lightlm} proposes a tailored Transformer-based recommender, which is effective and efficient for generative recommendations.  
Since \textbf{llamaRec~\cite{yue2023llamarec}} is a two-stage framework while \ourmodel is single-stage, we do not choose it as one baseline.

\end{enumerate*}

\subsubsection{Implementation details}
\label{sec:implementation_details}
We use T5-small from Huggingface\footnote{\url{https://huggingface.co/t5-small}} as our backbone in the main experiments. The encoder and decoder in this model both have 6 layers, each of which is an 8-headed attention layer. We find that after training, further fine-tuning the context-aware embeddings and the Recommendation LLM will result in better performance. In the ablation study ~\ref{sec:encoder}, we also investigate the impact of different backbones for Item LLM. Following ~\cite{li2023prompt}, we randomly sample a segment of no more than 21 items from a user's interaction history for each iteration which is denoted as '\ourmodel\_sampling' in Table~\ref{tb:main_result}. We also implement another training strategy where we traverse all the training data without sampling which is represented as '\ourmodel\_all'. We set the batch\_size for three datasets to 256 and the learning rate to 0.0005. The embedding dimension is set to 512. The dropout rate is 0.8 and the weight\_decay is 0.1 for Movies and Toys. The dropout rate is 0.7 for ML-1m. The warm\_up rate and the adam\_eps are set to 0.1 and 1e-6 for Movies and Toys, respectively. All methods are implemented using Pytorch with an AdamW optimizer. 
For the hyper-parameters of baselines, we use the values suggested by the original papers with carefully fine-tuning on the three datasets. We check the validation performance every epoch and adopt early-stop when the validation performance of R@10 doesn't improve for 20 consecutive times. 

\subsection{Effectiveness Comparison}
\label{sec:5.2}
The performance comparison is shown in Table~\ref{tb:main_result}. From the results, we can make the following observations: 1) \ourmodel exhibits significantly better performance than LLM-based recommendation baselines. Notably, the average Recall@10 improvements over the best results of LLM-based recommendation baselines are 46.8\% on ML-1m, 38.4\% on Movies, and 20.7\% on Toys, highlighting the effectiveness of our design. The potential reasons for the performance improvement are two-fold. First, the beam search decoding process will score  the whole LLM vocabulary where item tokens are a small proportion  while the item projection head in \ourmodel only scores items directly. Second, we propose a hierarchical LLM structure, where item information is encoded into context-aware vectors instead of the original natural language form. 2) \ourmodel consistently outperforms the best performance over baselines. The improvement on NDCG@20 is approximately 51.4\% on ML-1m, 11.3\% on Movies, and 7.9\% on Toys, which can be attributed to the power of LLMs. Notice that traditional ID-based and context-aware recommendation algorithms are still competitive. We will perform ablation study in Section~\ref{sec:component} to further study how \ourmodel works. 

\subsection{Efficiency Comparison}
\label{sec:5.3}
In this subsection, we analyze the inference efficiency of \ourmodel with LLM-based recommendation baselines. Since operations like obtaining context-aware embeddings and metrics computing and so on can be done off-line, we just measure the time between inputting the data to the model and obtaining the final recommendations as the inference time. The results are shown in Table~\ref{tb:inference_result}. Apparently, \ourmodel can achieve superior inference efficiency. The improvement over POD is approximately 99.71\% on ML-1m, 99.63\% on Movies, and 99.66\% on Toys. We contribute the efficiency improvement to the following reasons. First, we remove the most time-consuming part - beam search decoding. Second, the hierarchical LLM structure we propose to process long context can mitigate the redundant computation problem and reduce the computational costs. We also report the comparison of input length in Table~\ref{tb:length_result}. As can be seen, the advantage of \ourmodel is evident, as the improvement of input length is 75.2\%, 76.7\%, and 75.0\% over POD on three datasets, respectively.

\begin{table}[t]
\caption{Comparison of inference time (ms) for a batch of 32 users.}
\label{tb:inference_result}
\centering
    \begin{tabular}{l rrrr}
    \toprule
    \textbf{Datasets} & \textbf{ML-1m} & \textbf{Movies} & \textbf{Toys} \\ 
    \midrule
    P5   & 2,280 & 1,770 & 1,620 \\
    POD  & 2,170 & 1,400 & 1,490  \\ 
    \midrule
    \ourmodel & \textbf{6.13} & \textbf{5.14} & \textbf{5.03}\\
    \midrule
    Improvement & 99.71\% & 99.63\% & 99.66\% \\
    \bottomrule
    \end{tabular}
\end{table}

\begin{table}[t]
\caption{Comparison of average input length for a batch of 32 users.}
\label{tb:length_result}
\centering
    \begin{tabular}{l rrrr}
    \toprule
    \textbf{Datasets} & \textbf{ML-1m} & \textbf{Movies} & \textbf{Toys} \\ 
    \midrule
    P5   & 112.2 & 112.1 & 107.1 \\
    POD  & 85.0 & 89.1 & 83.5  \\ \midrule
    \ourmodel & \textbf{21} & \textbf{20.7} & \textbf{20.8}\\
    \midrule
    Improvement & 75.2\% & 76.7\% & 75.0\% \\
    \bottomrule
    \end{tabular}
\end{table}

\begin{figure*}[t]
    \centering
    \includegraphics[width=\linewidth]{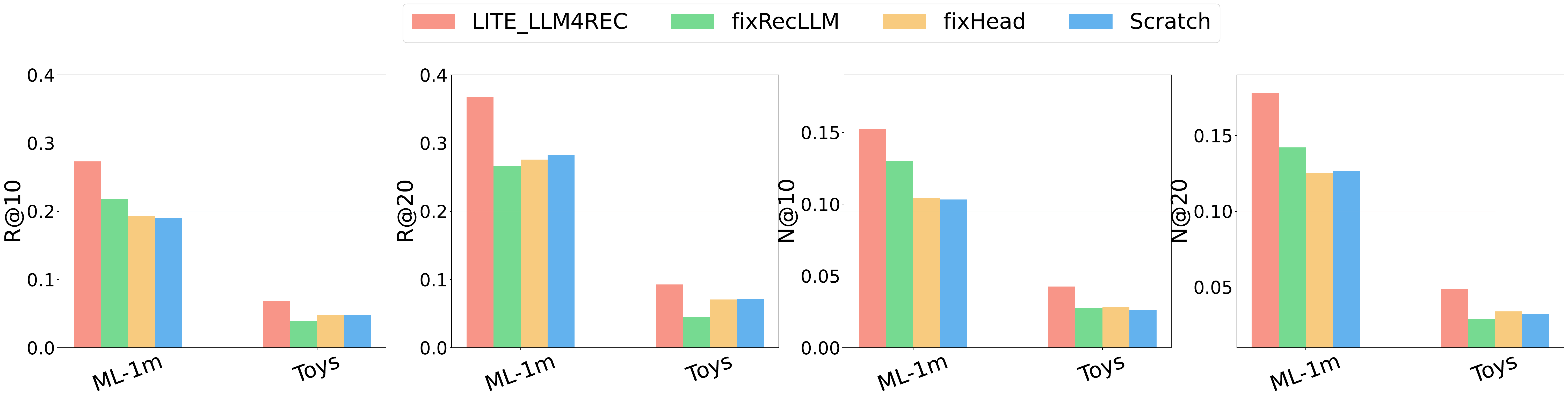}
    \caption{Ablation study results of R@10, R@20, N@10 and N@20 on ML-1m and Toys dataset.
    }
    \label{fig:ablation_study}
    \vspace{0.2cm}
\end{figure*}

\subsection{Top-N Recommendation}
\label{sec:5.4}
\begin{figure}[t]
    \centering
    \includegraphics[width=\linewidth]{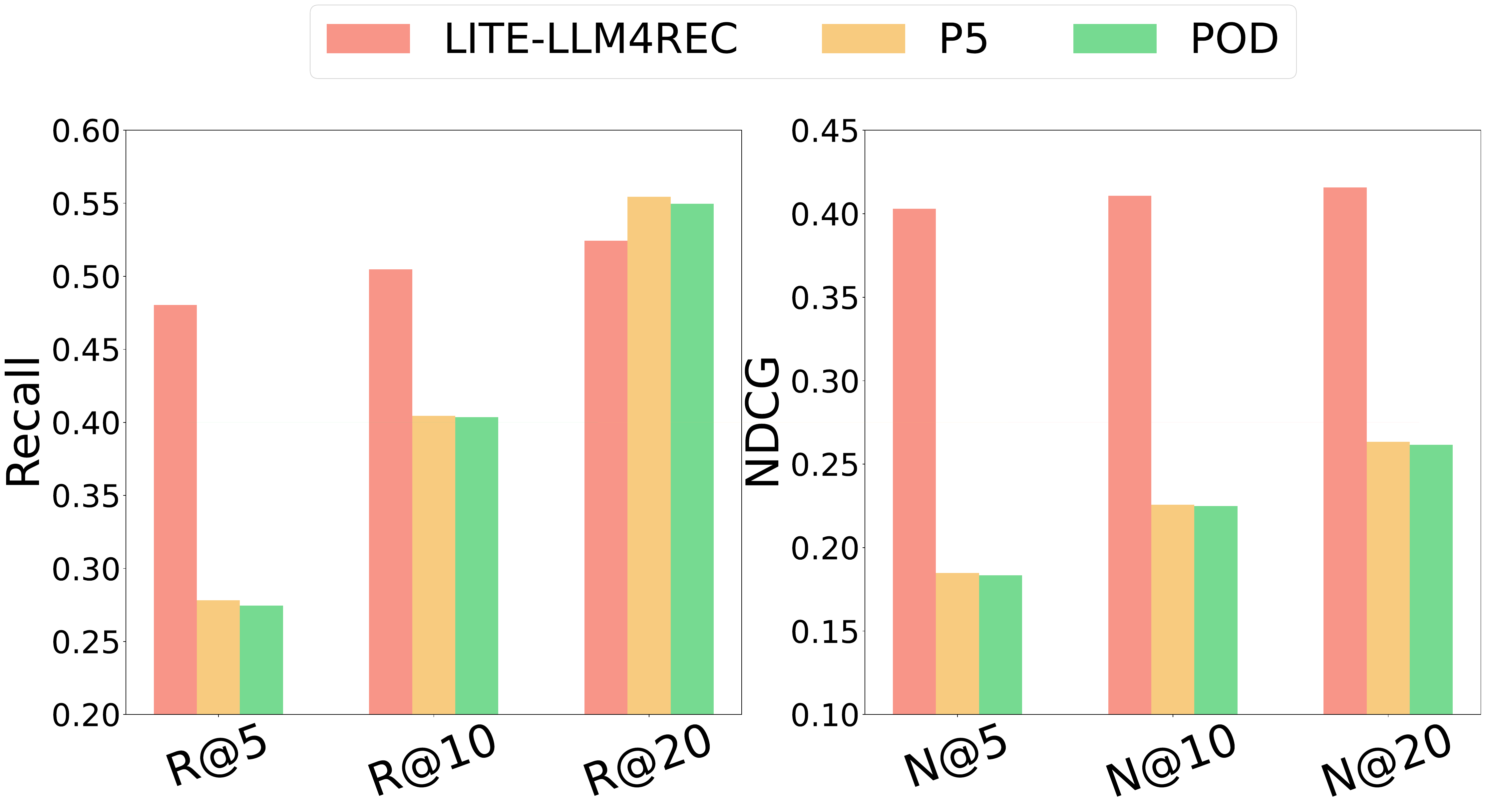}
    \caption{Results of Top-N recommendation on ML-1m. 
    }
    \label{fig:topn}
\end{figure}

Besides sequential recommendation, we also apply \ourmodel to Top-N recommendation task. We follow the setting of POD~\cite{li2023prompt}. We input one ground-truth item along with 99 negative items to the LLM and fine-tune the model to predict the ground-truth item. Finally, We test the model over the 99 negative examples. The results are reported in Figure~\ref{fig:topn}.

From the figure, we can find that \ourmodel can also perform well on this task. Our method can achieve much better performance on NDCG than P5~\cite{geng2022recommendation} and POD~\cite{li2023prompt}. This indicates that our method can greatly improve the ranking quality especially when candidates are given. For Recall, our method can perform better when $k$ is small. This is because the leave-one-out strategy we adopt where we only have one ground-truth item and our method can already rank the ground truth item in high position.

\subsection{Ablation Study}
\label{sec:5.5}

\subsubsection{Effectiveness of each component}
\label{sec:component}
In this section, we aim to analyse how different components influence the overall performance. We conduct experiments on ML-1m and Toys datasets with \ourmodel\_sampling strategy to assess each component. We design the following variants of our model: 
\begin{enumerate}
    \item fixT5: Recommendation LLM (the encoder of T5) is fixed. 
    \item fixHead: item projection head is fixed.
    \item Scratch: the parameters of the Recommendation LLM is randomly initialized instead of loading pre-trained weights.
\end{enumerate}

We demonstrate the results in Figure~\ref{fig:ablation_study}. From the results, we can have the following observations. 
First, each component in our framework contributes to the overall performance since fixing any one of the components results in the performance drop. 
Second, if we train a T5 from scratch instead of using the pre-trained weights for the recommendation LLM, the performance will drop either. This demonstrates the knowledge stored in LLM is of help to our recommendation task. 

\subsubsection{Impact of Item LLMs}
\label{sec:encoder}
In order to explore whether the performance gains come from the alignment between the Item LLM and the Recommendation LLM (both of them are T5), we conduct experiments with other Item LLM backbones like Bert~\cite{devlin2018bert}, Sbert~\cite{reimers2019sentence} and T5-base~\cite{raffel2020exploring}. For Bert, we use 'bert-base-uncased' version and take the output of the pooler layer as the sequence representation. For Sbert, we use 'all-MiniLM-L6-v2' version. Since the dimension of hidden state is not matched, one-layer MLP is adopted to transform the dimension. The results are reported in Table~\ref{tb:ab_encoders}. We can find that other Item LLMs can also achieve satisfying performance, which indicates that the influence of the backbones of Item LLMs is limited. 

\begin{table}[t]
\caption{Impact of Item LLMs}
\label{tb:ab_encoders}
\centering
    \begin{tabular}{l rrrr}
    \toprule
    ML-1m & R@10 & R@20 & N@10 & N@20\\ 
    \midrule
    \ourmodel  & 0.2733 & 0.3770 & 0.1518 & 0.1780 \\
    \midrule
    T5-base  & 0.2728 & 0.3742 & 0.1512 & 0.1767 \\
    Bert  & 0.2706 & 0.3749 & 0.1529 & 0.1793 \\
    Sbert    & 0.2668 & 0.3707 & 0.1486 & 0.1748 \\
    \bottomrule
    \end{tabular}
\end{table}


\section{Conclusion}
In this work, we propose \ourmodel, a simplified but effective LLM-based sequential recommendation model, which can achieve low-latency inference as well as better performance. Through experimental explorations, we find that the beam search decoding process is burdensome and unnecessary for sequential recommendation task. We propose to circumvent it and use an item projection head to acquire recommendations. Additionally, we find that existing item indexing methods will lead to high computational costs and redundant computations. To address the issue, we propose a novel hierarchical LLM structure to process the long context information of items efficiently while also enjoying the power of LLMs. These two designs effectively tackle the inference problem in existing LLM-based recommendation algorithms. Experiments conducted on three real-world datasets can demonstrate that \ourmodel can achieve superior inference efficiency while also improving the overall performance. In the future, we are interested in empowering our models with inductive learning ability. Also, we are interested in how the item indexing impacts the training of the model. Meanwhile, the size of the backbone of recommendation LLMs may affect the performance and we would investigate it as one future work.


\bibliographystyle{ACM-Reference-Format}
\bibliography{references}

\appendix









\end{document}